\documentclass[aps,pra,twocolumn,notitlepage,floatfix,nofootinbib,longbibliography]{revtex4-2}
\usepackage[latin9]{inputenc}
\usepackage{flafter}
\usepackage{verbatim}
\usepackage{cprotect}
\usepackage{amsmath}
\usepackage{amssymb}
\usepackage{graphicx}

\makeatletter

\DeclareTextSymbolDefault{\textquotedbl}{T1}

\usepackage{natbib}
\usepackage{amsfonts}
\usepackage{hyperref}
\usepackage{titlesec}
\titlespacing*{\subsection}{0pt}{*0.95}{*0.95}

\makeatother

\begin{document}
\title{Directional driving of vortex lines with oscillating magnetic field}
\author{A. E. Koshelev}
\affiliation{Department of Physics and Astronomy, University of Notre Dame, Notre
Dame, Indiana 46556, USA}
\date{\today }
\begin{abstract}
The possibility of driving vortex lines with an oscillating magnetic field
could be useful in many applications. For example, it can be used
for the removal of undesired trapped flux from contactless elements of
superconducting devices. We investigate the dynamics of vortex lines
in a superconducting film with a ratchet thickness profile driven by an oscillating
magnetic field applied parallel to the film. We numerically simulate
the dynamics of a single flux line modeled as an elastic string with
a variable length. Exploring the behavior for different frequencies
and amplitudes of the oscillating magnetic field, we find several
dynamic regimes. For moderate frequencies, the average velocity
is finite only within specific amplitude ranges. A notable
feature is the presence of extended velocity plateaus, which correspond
to regimes when the line moves by integer multiples of the spatial period
$w$ during integer multiples of the time period $T$.
The transitions to these plateau states are rather steep, especially
at low frequencies. The plateau at velocity $w/T$ dominates at intermediate
frequencies but vanishes at high frequencies. The onset field amplitude
of finite velocity nonmonotonically depends on the frequency and passes
through a minimum at a certain frequency value. At low frequencies,
the velocity exceeds $w/T$ and progressively increases with the amplitude.
These findings provide valuable insights into the dynamic behavior
of vortex lines driven by oscillating magnetic field in patterned
superconducting films, offering potential pathways for controlling
the magnetic flux in superconducting devices.
\end{abstract}
\maketitle

\section{Introduction\protect\label{sec:Introduction}}

Trapped magnetic flux poses a major issue for many superconducting
devices \citep{Tanaka2001,Polyakov2007,Narayana2009,Jackman2017,Jabbari2022}.
Trapped Abrikosov vortices increase dissipation, generate additional
noise, and suppress the critical currents of Josephson junctions. 
These effects result in undesirable variations in the operational parameters of superconducting electronic circuits. 
Therefore, efficient schemes for elimination of trapped vortices would
be very beneficial for future applications of superconducting electronics. However, the controlled manipulation of vortex lines in
type-II superconductors remains a long-standing challenge.

A promising concept is to drive out the trapped vortex lines with
an oscillating current using ratchet potentials \citep{LeeNature1999,WambaughPhysRevLett99,ZhuPhysRevB03,LuPhysRevB07,ShklovskijPhysRevB11,ReichhardtPhysRevB15}.
This theoretical proposal has been demonstrated for several implementations
of asymmetric pinning potential in different superconducting materials.
Specifically, the rectified voltage caused by the directed vortex
motion induced by an oscillating current was observed in a niobium
film deposited on an array of aligned nickel triangular dots\citep{Villegas2003,VillegasPhysRevB05},
in aluminum films with square array of hole pairs with different sizes\citep{VondelPhysRevLett05,SouzaSilvaPhysRevB06,SouzaSilvaNature2006},
in lead films with nanoengineered asymmetric antidots\citep{JinPhysRevB10}, 
in epitaxial Nb films decorated with ferromagnetic Co nanostripes\citep{DobrovolskiyPhysRevApplied.13.024012},
in MoGe films with conformal-mapped nanoholes having a density gradient\citep{LyuNatComm2021}.
Directed motion of vortices in niobium films with an asymmetric array
of pinning sites was visualized by Lorentz microscopy \citep{TogawaPhysRevLett05}.
Another manifestation of asymmetric pinning potentials is the finite
difference between critical currents for two opposite directions.
Such asymmetry was indeed detected 
in Nb \citep{DobrovolskiyAPL2015} and 
YBa$_{2}$Cu$_{3}$O$_{7-\delta}$ (YBCO) \citep{JonesACSAMI2020} films with a sawtooth thickness profile and 
in YBCO films with array of asymmetric antidotes\citep{PalauPhysRevB.85.012502}.
Extensive molecular-dynamics
\citep{OlsonPhysRevLett.87.177002,ZhuPhysRevB03,ZhuPhysRevLett.92.180602,GillijnsPhysRevLett07,ReichhardtPhysRevB15}
and time-dependent Ginzburg-Landau \citep{LyuNatComm2021} simulations
of vortex dynamics revealed many insights into the dynamic response
for different realizations of ratchet potentials. These simulations
quantitatively characterized the sensitivity of rectified vortex transport
to key parameters including frequency of AC drive, vortex density,
thermal noise, the strength of random pinning, and the ratchet shape. These
simulations and experiments set a fundamental basis for elaboration
of efficient schemes for control and manipulation of magnetic flux
in superconducting devices.

In all existing theoretical proposals and experimental realizations
of vortex ratchets, an oscillating electric current has been used
as a driving force imposing directional motion of vortices through
ratchet potentials. In some situations, however, passing of current
through circuit elements may be either not practical or impossible.
An attractive alternative is to use instead an oscillating in-plane
magnetic field. Such a magnetic field induces tilting of vortex lines.
In ratchet profiles without top/bottom symmetry such tilting may induce
directional motion of the line due to the rectification of the top
tip motion. This would permit flux removal from contactless elements,
such as stand-alone stripline resonators. In this paper, we explore
the feasibility of such approach with numerical simulations of a single
vortex line driven by an oscillating magnetic field in a superconducting
film with 
an asymmetric pinning potential. 
As a specific realization of such an asymmetric potential, we select a sawtooth modulation of film thickness, 
as in the original theoretical suggestion \citep{LeeNature1999}.
This is the simplest ratchet potential simultaneously breaking both left/right and top/bottom symmetries.
Previously, such asymmetric thickness modulation was experimentally realized  
in Nb \citep{DobrovolskiyHuthAPL2015,DobrovolskiyAPL2015} and  YBCO \citep{JonesACSAMI2020}.
The effect discussed in this paper may be realized for other versions of the asymmetric ratchet potential.

We model the line as an elastic
string with variable length and explore dynamic behavior for different
frequencies and amplitudes of the oscillating magnetic field. To demonstrate
the feasibility of this approach, we investigate in detail a specific
realization of ratchet potential for which the screening current induced
by the oscillating magnetic field is given by a simple local expression.
We found that the vortex line indeed can be driven by an oscillating
magnetic field and our study revealed a rich and nontrivial dynamic
behavior.

The paper is organized as follows. In Sec.~\ref{sec:Model}, we describe
the model we use to simulate the vortex line driven by the oscillating
in-plane magnetic field. In Sec\@.~\ref{sec:Results}, we present
and discuss the simulation results using the numerical procedure described
in Appendix \ref{app:NumProc}. We explore the stability range of
the line configuration in a static in-plane magnetic field and investigate
the dynamic response of the line at different amplitudes and frequencies
allowing us to establish several distinct dynamic regimes. Finally,
we summarize in Sec.~\ref{sec:Summary}.

\section{Model\protect\label{sec:Model}}

\begin{figure}
\includegraphics[width=3in]{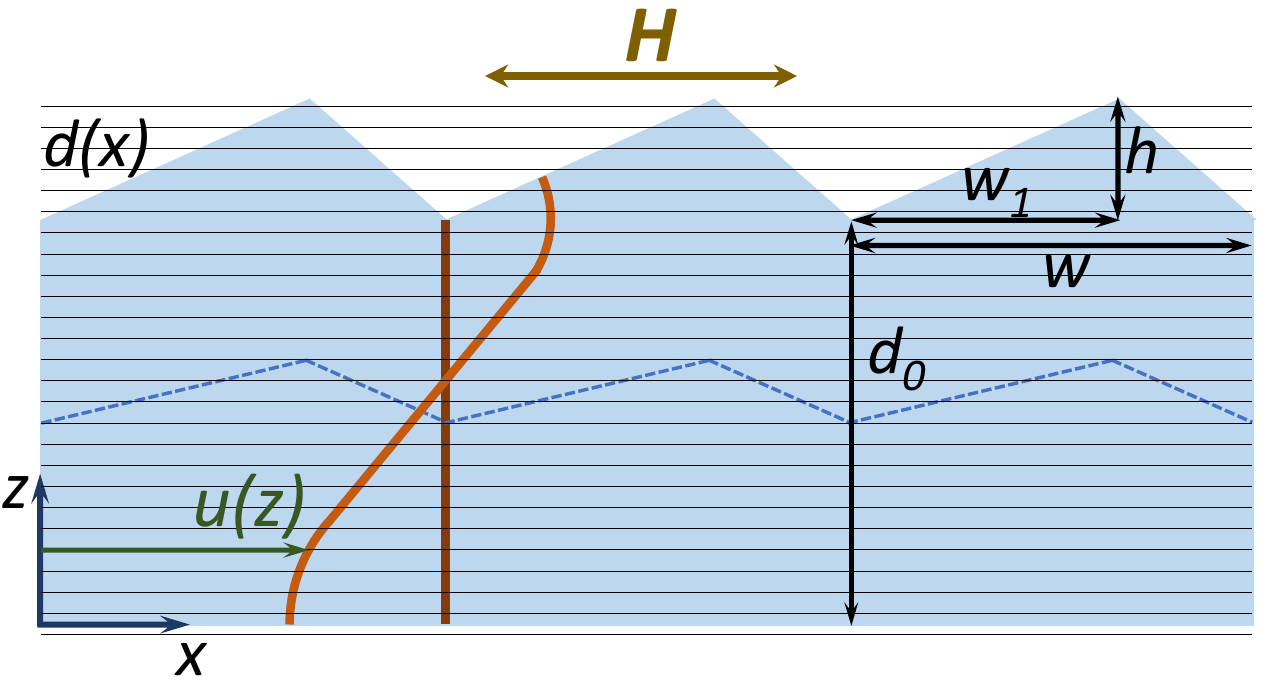}
\caption{Illustration of a deformed vortex line inside a superconducting film
with a ratchet thickness profile. The notations for the geometrical
ratchet parameters are shown. The short-dash line shows the line of
zero screening current for the local approximation in Eq.\ \eqref{eq:Current}. The horizontal lines illustrate the numerical
mesh. }
\label{fig:VortLineRatchet}
\end{figure}
We consider a superconducting film with a nonuniform thickness $d(x)$ 
forming an asymmetric ratchet profile, 
\begin{equation}
d(x)=
\begin{cases}
	d_{0}+\frac{h}{w_{1}}x, & \mathrm{for}\:0\!<\!x\!<\!w_{1}\\
	d_{0}\!+\frac{h}{w_{2}}\left(w\!-\!x\right), & \mathrm{for}\:w_{1}\!<\!x<\!w
\end{cases},
\label{eq:RatchetProfile}
\end{equation}
with $w\!=\!w_{1}\!+w_{2}$ being the ratchet period. The film occupies the region $0\!<\!z\!<\!d(x)$. 
The geometry
of the patterned film and the definitions of the parameters are illustrated
in Fig.~\ref{fig:VortLineRatchet}. Having in mind the problem of
removal of accidental trapped flux with very low density, we will
study dynamics of a single vortex line. Initially, this line is trapped
in the minimum of this ratchet potential, as illustrated by the vertical
dark brown line in Fig.~\ref{fig:VortLineRatchet}. An in-plane magnetic
field applied along the $x$ axis generates a screening supercurrent
along the $y$ axis which tilts the vortex line. In general, the calculation
of the distribution of the supercurrent for a patterned film is a
nontrivial problem. Here we focus on the simplest case when the thickness
is much smaller than the London penetration depth, $d(x)\!\ll\!\lambda$
and thickness variation is smooth, $\mathrm{d}d/\mathrm{d}x\ll1$, $w_1,w_2\gg \lambda$.
In this case, a nonuniform supercurrent $j(x,z)$ generated by the
in-plane magnetic field $H$ is given by the local approximation in the
thin-film limit,
\begin{equation}
j(x,z)=\frac{c}{4\pi\lambda^{2}}H\left(z\!-\frac{d(x)}{2}\right).\label{eq:Current}
\end{equation}
On the other hand, we assume that the coherence length $\xi$ is smaller
than the thickness. We consider a curved vortex line described by
a displacement function $u(z)$, as illustrated in Fig.~\ref{fig:VortLineRatchet}.
We model the line as an elastic string. Even though such a description
is not exact due to the nonlocality of the line energy \citep{Brandt1995},
it is sufficient for understanding qualitative behavior and has been
employed before in several simulation studies\citep{OtterloPhysRevLett98,Koshelev2011,PleimlingPhysRevB.84.174509,DobramyslPRE2014}.
The energy of the arbitrarily deformed line inside an isotropic superconductor
can be written as 
\begin{align}
\mathcal{E} & =\epsilon_{1}\int_{0}^{d(u_{d})}\!dz\,\sqrt{1\!+\!\left(\frac{\mathrm{d}u}{\mathrm{d}z}\right)^{2}}\nonumber \\
- & \frac{\Phi_{0}}{c}\int_{0}^{d(u_{d})}\!dz\int_{0}^{u(z)}\!dx\,j(x,z)\label{eq:LineEnergy}
\end{align}
where $\epsilon_{1}\simeq\frac{\Phi_{0}^{2}}{(4\pi\lambda)^{2}}\ln\frac{d_{0}}{\xi}$
is the line tension (with logarithmic accuracy) and $u_{d}\!=\!u\left[d(u_{d})\right]$
is the displacement of the top tip of the vortex line. The length
of the line in the $z$ direction varies depending on this displacement.
In the second term, we can formally set $j(x,z)\!=\!0$ in the regions
outside the superconductor. For the current in Eq.~\eqref{eq:Current},
the second term in Eq.~\eqref{eq:LineEnergy} becomes 
\[
-\frac{\Phi_{0}H}{4\pi\lambda^{2}}\int_{0}^{d(u_{d})}\!\!dz\int_{0}^{u(z)}\!\!dx\!\left(z\!-\frac{d(x)}{2}\right).
\]
By varying the energy with respect to the displacement, we find the
dynamic equation 
\begin{equation}
\eta\frac{\mathrm{d}u}{\mathrm{d}t}=\epsilon_{1}\frac{\mathrm{d}}{\mathrm{d}z}\frac{1}{\sqrt{1\!+\!\left(\frac{\mathrm{d}u}{\mathrm{d}z}\right)^{2}}}\frac{\mathrm{d}u}{\mathrm{d}z}+\frac{\Phi_{0}H}{4\pi\lambda^{2}}\left(z\!-\frac{d(u)}{2}\right),\label{eq:VLdyn}
\end{equation}
with $\eta$ being the vortex-line viscosity coefficient and the boundary
conditions\begin{subequations}
\begin{align}
\frac{\mathrm{d}u}{\mathrm{d}z} & =0,\,\mathrm{for}\,z\,=\,0,\label{eq:BC0}\\
\frac{\mathrm{d}u}{\mathrm{d}z} & =-\frac{\mathrm{d}d}{\mathrm{d}x},\,\mathrm{for}\,z\,=\,d(u_{d}),\label{eq:BCd}
\end{align}
\end{subequations}implying that the line is always locally oriented
perpendicular to the surface. We will study the dynamic response to
the oscillating magnetic field, $H(t)=H_{0}\cos\left(\omega t\right)$. 

We note that, as our goal is to demonstrate the feasibility of directional drive with the oscillating magnetic field,
we use the simplest elastic-line model described by Eqs.\ \eqref{eq:VLdyn}, \eqref{eq:BC0}, and \eqref{eq:BCd}.  
Admittedly, this model does not fully realistically describe a vortex line in patterned films. For example, it does not take into account long-range interactions with the ratchet profile. A realistic description of vortex-line dynamics can be achieved using the time-dependent Ginzburg-Landau model which is much more computationally demanding. Our model also neglects several physical properties that are potentially relevant in real materials, such as vortex pinning, thermal noise, and intervortex interactions.  We can note that the strength of thermal noise quantified by  the Ginzburg number \cite{LarkinVarlBook2009} strongly varies for different superconducting materials. In particular, in the target material for this study, niobium, the thermal noise usually can be neglected due to a very small Ginzburg number ($\sim 7\cdot 10^{-12}$ in clean case). However, for other superconducting films with smaller coherence length and larger London penetration depth, the thermal fluctuations may indeed be essential in the vicinity of the transition temperature.  
The effect discussed here may be suppressed by strong vortex pinning. Pinning is not essential when the critical current it produces is smaller than both the ratchet-induced critical current and the screening current generated by the oscillating magnetic field. The employed single-line approximation is justified at small magnetic fields when the intervortex interaction forces are smaller than forces due to ratchet slope. 

For numerical calculations, we introduce the reduced coordinates,
time, and magnetic field as $\tilde{z}=z/d_{0}$, $\tilde{u}=u/d_{0},$
\begin{align}
\tilde{t} & =\frac{\epsilon_{1}}{\eta d_{0}^{2}}t,\label{eq:Reducedt}\\
\tilde{H} & = H/H_{\mathrm{u}}=\frac{\Phi_{0}d_{0}^{2}}{4\pi\lambda^{2}\epsilon_{1}}H.\label{eq:ReducedH}
\end{align}
Correspondingly, the units of angular frequency and velocity are
\begin{align}
\omega_{\mathrm{u}} & =\frac{\epsilon_{1}}{\eta d_{0}^{2}},\label{eq:frUnit}\\
v_{\mathrm{u}} & =\frac{\epsilon_{1}}{\eta d_{0}}.\label{eq:velUnit}
\end{align}
For fixed $z$-axis magnetic field $B_{z}$, the electric field in
terms of the reduced velocity $\tilde{v}$ can be expressed as
\begin{equation}
E=\tilde{v}\frac{\epsilon_{1}}{\eta d_{0}c}B_{z}.\label{eq:ElField}
\end{equation}
We evaluate the field, frequency, and velocity scales for several materials below, in Section \ref{Sec:ExpParams}.  

In the reduced variables, the reduced energy $\tilde{\mathcal{E}}\!=\!\mathcal{E}/\epsilon_{1}d_{0}$
following from Eq.~\eqref{eq:LineEnergy} can be written as
\begin{equation}
\tilde{\mathcal{E}}\!=\!\!\int\limits_{0}^{\tilde{d}(\tilde{u}_{d})}\!\!d\tilde{z}\left[\,\sqrt{1\!+\!\left(\frac{\mathrm{d}\tilde{u}}{\mathrm{d}\tilde{z}}\right)^{2}}\!-\!\tilde{H}\!\int\limits_{0}^{\tilde{u}(\tilde{z})}\!d\tilde{x}\!\left(\tilde{z}\!-\frac{\tilde{d}(x)}{2}\right)\right],\label{eq:LineEnergyR}
\end{equation}
and the dynamic equation takes the form 
\begin{equation}
\frac{\mathrm{d}\tilde{u}}{\mathrm{d}\tilde{t}}=\frac{\mathrm{d}}{\mathrm{d}\tilde{z}}\frac{1}{\sqrt{1\!+\!\left(\frac{\mathrm{d}\tilde{u}}{\mathrm{d}\tilde{z}}\right)^{2}}}\frac{\mathrm{d}\tilde{u}}{\mathrm{d}\tilde{z}}+\!\tilde{H}\left(\tilde{z}\!-\frac{\tilde{d}(\tilde{u})}{2}\right).\label{eq:VLdynRed}
\end{equation}
We proceed with presentation and discussion of results obtained from
numerical simulations of this equation with boundary conditions in
Eqs.~\eqref{eq:BC0} and \eqref{eq:BCd}.

\section{Results and discussion\protect\label{sec:Results}}

We explore the response of a single vortex line to the oscillating
in-plane magnetic field by numerical simulations of Eq\@.~\eqref{eq:VLdynRed}
for a wide range of amplitudes and frequencies. The numerical
procedure is described in Appendix \ref{app:NumProc}. For detailed investigation,
we select a representative realization of the ratchet profile with
$w\!=\!8d_{0}$, $h\!=\!0.5d_{0}$ and $w_{1}\!=\!6d_{0}$. A large
lateral size is selected to facilitate applicability of the local approximation
for the current profile in Eq.~\eqref{eq:Current}. In the presentations
of the numerical results, we will continue using the reduced units
in Eqs\@.~\eqref{eq:Reducedt} and \eqref{eq:ReducedH} but will
omit the marker \textquotedbl$\sim$\textquotedbl to simplify
the presentation.

\begin{figure}
\includegraphics[width=3.4in]{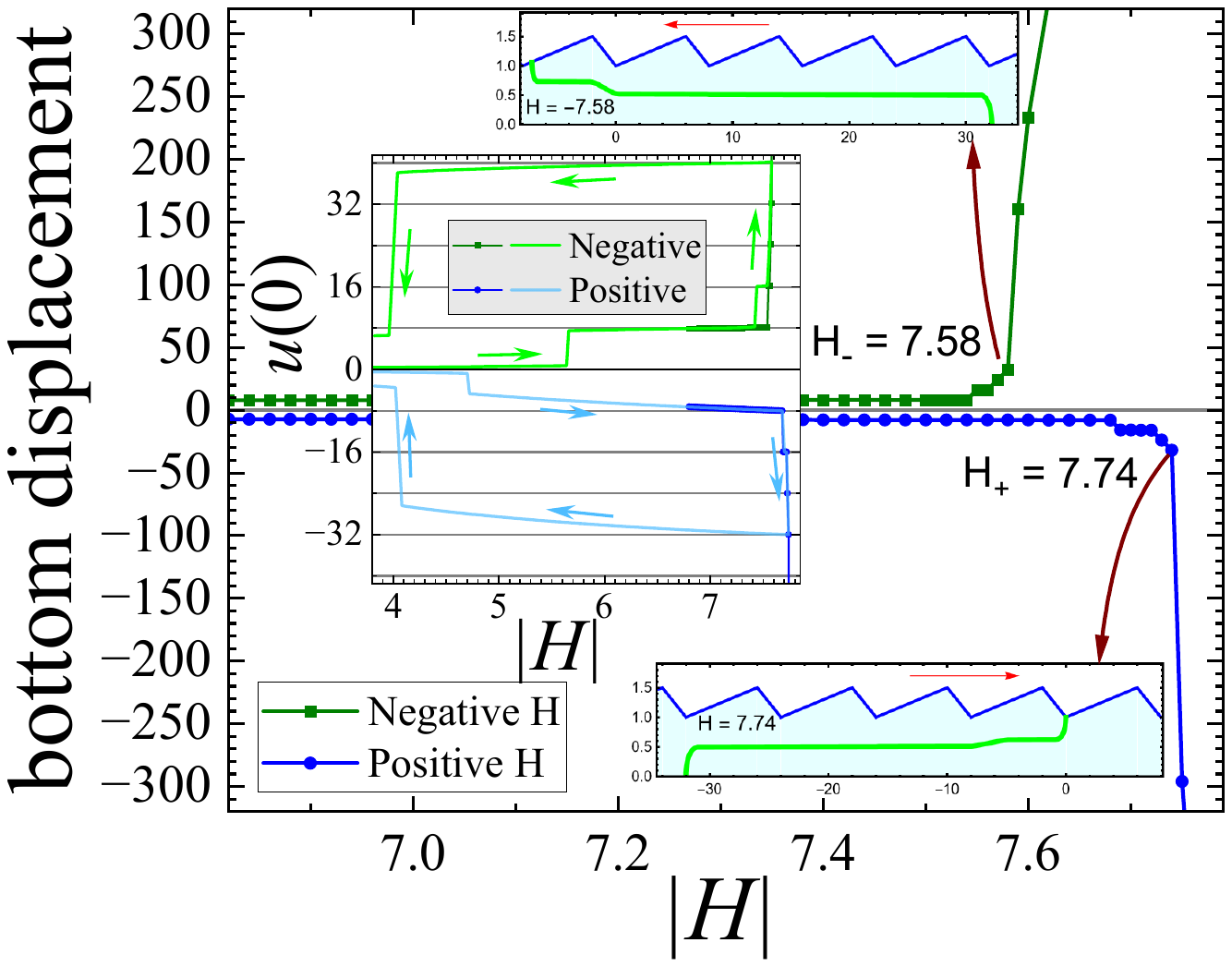} \caption{\emph{Main plot}: The displacement of the bottom tip over a long time
(2000 in reduced units) for two directions of applied static in-plane
magnetic field. The onset of the rapid increase marks the instability
of the static line configuration. The images show the last stable
configurations for two field directions. Note that, for clarity, we
set the horizontal scale five times smaller than the vertical scale
in all configuration pictures here and below. \emph{The inset} zooms
into the range of the stable static configuration and shows a hysteretic
behavior of the displacement when the magnetic field is ramped down
from the maximum static-stability value.}
\label{Fig:umaxH}
\end{figure}

\subsection{Stability of static configuration.}

 Only an oscillating magnetic field with sufficiently high amplitude
is capable of driving the vortex line in one direction. The relevant
field scale is the magnetic field at which the static line configuration
becomes unstable. We start with the evaluation of the stability field
for a uniform film with thickness $d_{0}$ smaller than the London
penetration depth $\lambda$. The screening supercurrent induced by
the in-plane magnetic field in Eq.~\eqref{eq:Current} tilts the
vortex line, and the static configuration of the line results from
a balance between this tilting force and the line tension. At sufficiently
high magnetic field, the line is composed of a segment of in-plane
vortex located in the middle of the film and two kinks connecting
it with the top and bottom surfaces. This configuration is stable
until the force acting on the kink from the screening current 
\[
f_{k}\!=\frac{\Phi_{0}}{c}\int_{d_{0}/2}^{d_{0}}\!j(z)dz\!
=\frac{\Phi_{0}H}{4\pi\lambda^{2}}\int_{0}^{d_{0}/2}\!\!zdz\!
=\frac{\Phi_{0}H}{4\pi\lambda^{2}}\frac{d_{0}^{2}}{8}
\]
is smaller than the line tension $\varepsilon_{1}$. The corresponding
instability field 
\begin{equation}
H_{\mathrm{inst}}=\frac{32\pi\lambda^{2}\varepsilon_{1}}{\Phi_{0}d^{2}}\approx\frac{2}{\pi}\frac{\Phi_{0}}{d_{0}^{2}}\ln\frac{d_{0}}{\xi}
\label{eq:Hinst}
\end{equation}
coincides with the in-plane lower critical field $H_{c1}$ for a film
thinner than the London penetration depth \citep{Abrikosov1988,StejicPhysRevB.49.1274}.
For the reduced instability field defined in Eq.~\eqref{eq:ReducedH},
we have $\tilde{H}_{\mathrm{inst}}\!=\!8$. This instability field
for the minimal thickness provides a rough estimate for the magnetic-field
scale at which DC drive may be expected for a film with a ratchet
profile. 

A second important in-plane field scale is the instability field of the Meissner state commonly referred to as the superheating field, $H_{\mathrm{sh}}$. The magnetic field exceeding this threshold will generate vortices at the surface. Analysis of this problem for bulk superconductors in the limit of a large Ginzburg-Landau parameter $\kappa$  \cite{KramerPhysLett1967, *KramerPhysRev.170.475, ChapmanSIAMJApplMath1995} demonstrates that the Meissner state becomes unstable when the reduced superconducting momentum $Q\!=\!\xi\left|\boldsymbol{\bigtriangledown}\phi\!-\frac{2\pi}{\Phi_{0}}\boldsymbol{A}\right|$ at the boundary exceeds $1/\sqrt{3}$. For a uniform film with thickness $d_{0}$ in the limit $\xi\ll d_{0}\ll\lambda$  this criterion yields
\begin{equation}
H_{\mathrm{sh}}	=\frac{\Phi_{0}}{\sqrt{3}\pi\xi d_{0}}.
\label{eq:Hsh}
\end{equation}
This field sets the upper limit for the amplitude of AC magnetic field that can be used to drive $z$-axis vortices. 

\begin{figure*}[t]
\includegraphics[width=5.8in]{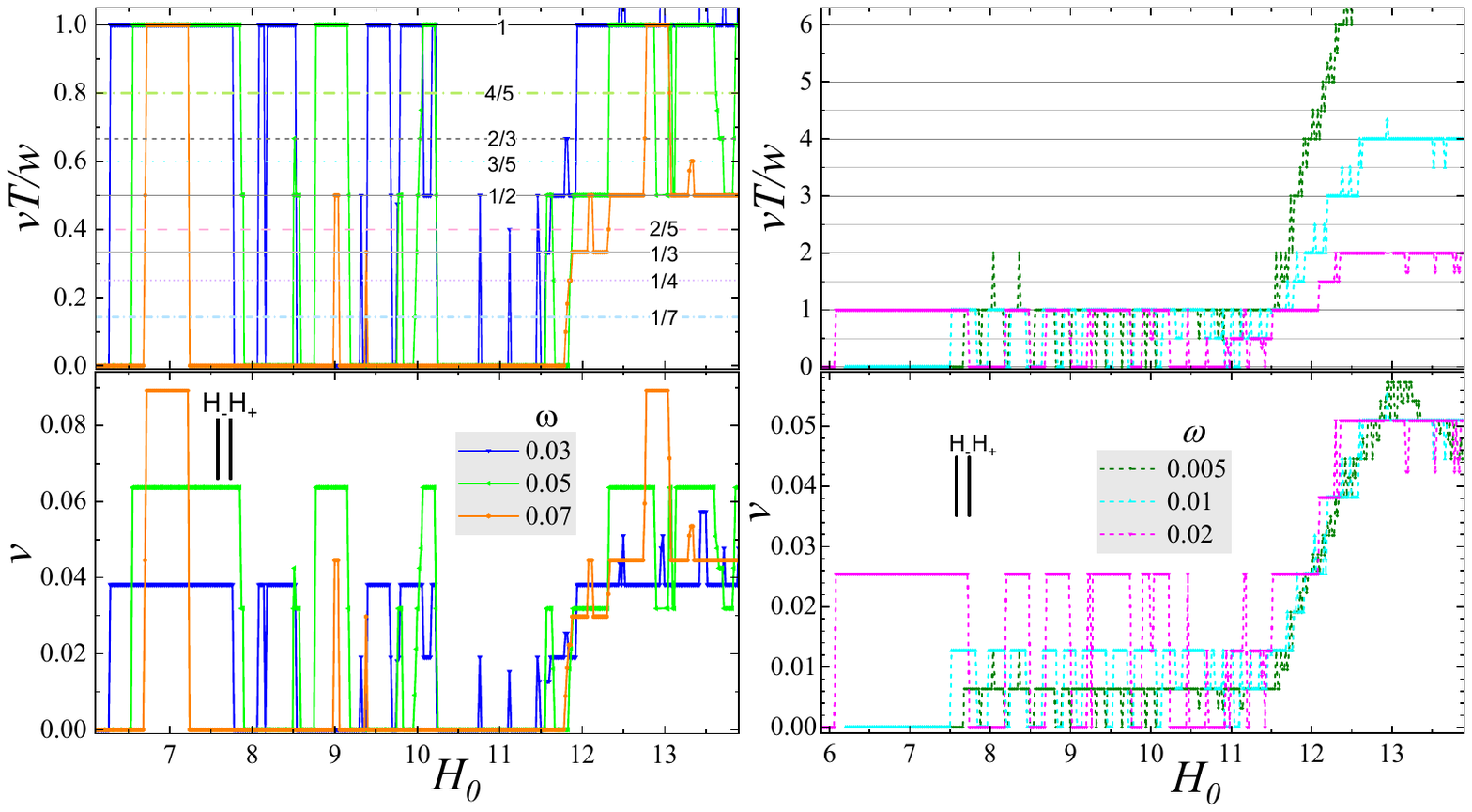}\caption{Dependences of the average velocity $v$ on the amplitude of the in-plane
AC field $H_{0}$ for several representative frequencies $\omega$. The units of the field amplitude and angular frequency are defined in Eqs.\ \eqref{eq:ReducedH} and \eqref{eq:frUnit}, respectively.
The bottom plots shows the velocity in units of of $v_{\mathrm{u}}$,
Eq.~\eqref{eq:velUnit}. For reference, we also show the values of
static instability fields, $H_{+}$ and $H_{-}$, see Fig.~\ref{Fig:umaxH}.
The top plots shows velocities normalized to $w/T$, where $w$ is
the ratchet period and $T=2\pi/\omega$ is the period of the AC drive.
The horizontal lines show the rational-number levels $n_{w}/n_{T}$
corresponding to the realized regimes of motion when the line moves
$n_{w}$ ratchet periods in $n_{T}$ time periods.}
\label{Fig:velHSelected}
\end{figure*}

For the film with the used ratchet thickness profile, we investigate
the stability of the static configuration numerically. The key qualitative
feature is that for an asymmetric ratchet, the instability fields
are different for two field directions. We start the simulation with
a vortex line oriented vertically and apply the static in-plane magnetic
field. The top tip of the line remains pinned in the minimum of the
ratchet profile and the bottom tip moves. At small fields, the line
reaches a stable configuration which does not change with time anymore. At a sufficiently large field, this configuration acquires the
shape of a long segment of the in-plane vortex and two kinks, see the
line configuration images in Fig.~\ref{Fig:umaxH}. We can also notice
that the bottom kink in the static critical configuration is located
beneath the ratchet dip, where the field-induced in-plane current
is minimal. Above a certain field the static configuration becomes
unstable and the bottom kink moves indefinitely increasing the length
of the in-plane vortex segment.

The main plot in Fig.~\ref{Fig:umaxH} shows displacements of the
bottom tip after a rather long simulation time equal to 2000 time
units for two field directions. We see that above certain field these
displacements increase abruptly marking the onset of instability.
Above the instability field, the bottom displacement moves with constant
speed. It is finite in the plot only because of a finite simulation
time. We also show the last stable line configurations at the instabilities.
These simulations allow us to estimate the reduced instability fields
as $H_{+}\!=7.74$ and $H_{-}\!=7.58$ for the positive and negative
filed direction, respectively. As expected, both fields are somewhat
smaller than the instability field for a uniform film with minimum
thickness $d_{0}$, which is equal to $8$ in the reduced units. The
inset in Fig.~\ref{Fig:umaxH} shows in better detail the behavior
inside the stable region. We can see that at large field amplitudes
the bottom displacement advances in discrete steps by jumping between
the locations under the ratchet dip. We also demonstrate that the
behavior is hysteretic: when the magnetic field is decreased from
the maximum static-stability value, the displacement does not follow
the same curve but remains pinned close to the maximum value within
rather a wide field range (down to $|H|=4.04$ and $4.08$ for the
negative and positive direction, respectively). This hysteretic behavior
has important implications for the dynamic response, especially at
low frequencies.

\subsection{Dynamic response}

\begin{figure}[t]
\includegraphics[width=3.4in]{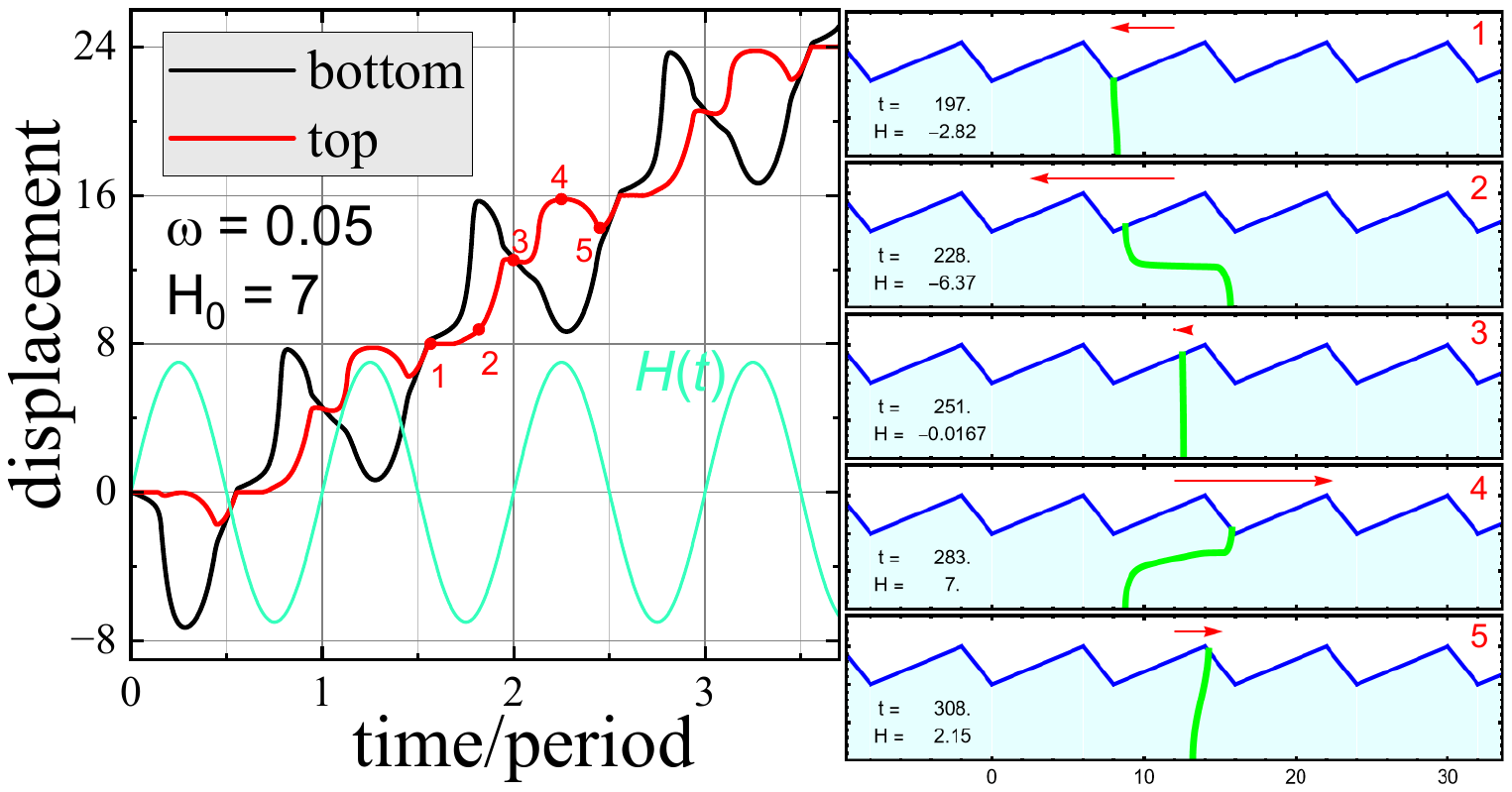}
\caption{The displacements of the bottom and top tips of the vortex vs time
for $\omega=0.05$ and $H_{0}=7$. For these parameters, the basic
regime of motion is realized when the line moves at one spatial period
$w$ during one time period $T$. For reference, the time dependence
of the magnetic field is also shown. The images on the right show
the line configurations at five times marked in the plot. Note that
in these images the horizontal scale is five times smaller than the
vertical one. The animation of the line dynamics for these parameters
can be found in the clip \ref{video1}.
}
\label{Fig:DsplTimeCnffr025H8}
\end{figure}
\begin{video}[h]
	\includegraphics[width=3.in]{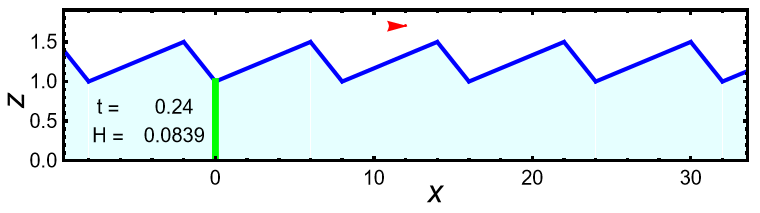}
	\setfloatlink{https://drive.google.com/file/d/155ogiFbAh1AxhKUx7_e212DYln5iuCWy/view?usp=drive_link}
	\caption{Animation of vortex line dynamics in the film with ratchet thickness profile for $\omega\! =\! 0.05$ and $H_0\!=\!7$.
	\label{video1}
	}
\end{video}
We proceed with investigation of the vortex-line dynamics in the
oscillating in-plane magnetic field based on numerical simulations
of Eq.~\eqref{eq:VLdynRed} with the boundary conditions in Eqs.~\eqref{eq:BC0}
and \eqref{eq:BCd}. We studied dynamics response for many reduced
frequencies and for the reduced amplitudes below $H_{0}\!=\!14$. For
our ratchet geometry, extended regions of directional motion for the
studied amplitude range appear at frequencies below $0.09\,\omega_{\mathrm{u}}$.

Figure \ref{Fig:velHSelected} shows the dependences of the average
DC velocity of the vortex line on the reduced amplitude of the AC
drive $H_{0}$ for several select frequencies in units of $\omega_{\mathrm{u}}$,
Eq.~\eqref{eq:frUnit}. The lower plots are for the velocity in units
of $v_{\mathrm{u}}$, Eq.~\eqref{eq:velUnit}, which is the same
for all frequencies. The left column shows plots for intermediate
frequency range from 0.03 to 0.07 and the right column shows plots
for low frequencies below 0.02. For reference, we also show the static
instability fields, $H_{+}$ and $H_{-}$, see Fig.~\ref{Fig:umaxH}.
We see that for every frequency in the studied range the velocity
is finite only within some amplitude ranges. Interestingly, for frequencies
above 0.02, the average velocity becomes finite above the onset field
amplitude, which is somewhat smaller than $H_{-}$. A salient feature
of these velocity dependences is the presence of extended plateaus,
corresponding to regimes when the line moves integer $n_{w}$ spatial
periods $w$ during integer $n_{T}$ time periods $T\!=\!2\pi/\omega$.
These plateaus reflect a matching effect between the spatial periodicity
of the ratchet potential and the temporal periodicity of the AC drive.
Similar velocity plateaus were observed for a two-dimensional vortex
lattice driven by the oscillating Lorentz force through a square array
of asymmetric pinning sites at the matching magnetic field when the
vortex and pin densities are equal \citep{ZhuPhysRevB03}. For a clearer
display of these plateau features, we show in the upper plots the velocity
normalized by the ratio of the ratchet period $w$ and the AC drive period
$T$ and display several levels $n_{w}/n_{T}$ corresponding to realized
dynamic states. We see that the most pronounced plateaus in the intermediate
frequency range are at the velocities $w/T$, $w/2T$, and $w/3T$
and the plateau width rapidly decreases with the denominator. The
transitions to these states are abrupt, especially at amplitudes below
$10$. 

\begin{figure}
\includegraphics[width=3.45in]{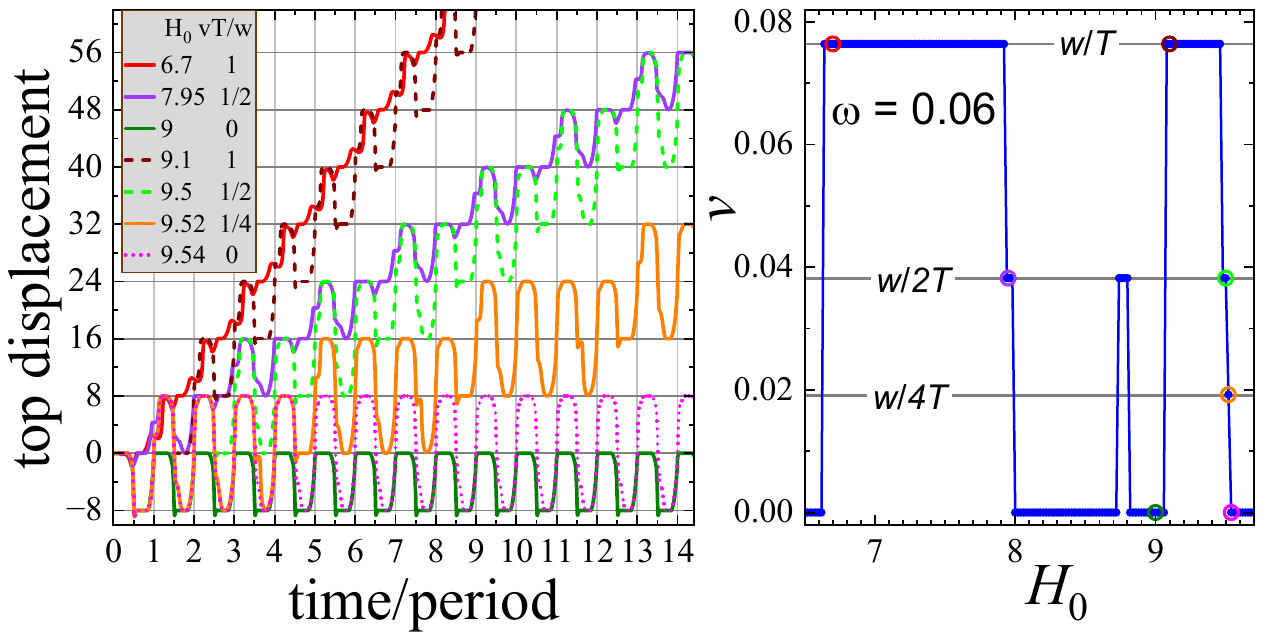}
\caption{Series of the time dependences of the top-tip displacement for the
reduced frequency of $0.06$ and several magnetic-field amplitudes
which represent different regimes of motion corresponding to different
rational numbers of the parameter $vT/w$. The presented field amplitudes
are marked in the velocity-amplitude plot shown on the right. The
animations of the line dynamics for $H_{0}=9.1$, $9.5$, and $9.52$
can be found in the clip \ref{video2}.
}
\label{Fig:udtfr0_06}
\end{figure}
\begin{video}[h]
	\includegraphics[width=3.in]{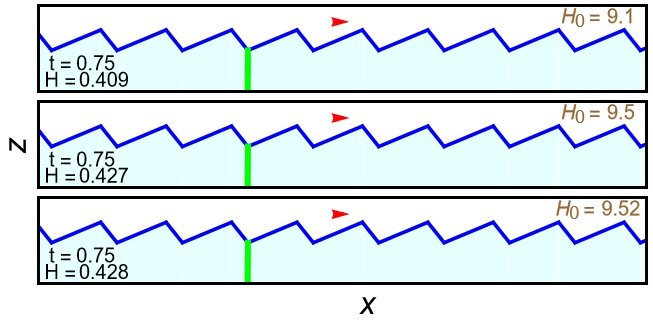}
	\setfloatlink{https://drive.google.com/file/d/164ZQMGg2hKFO9aoT7w6iZ6_qVldAfpgp/view?usp=drive_link}
	\caption{Animations of a vortex line dynamics in the film with ratchet thickness profile for $\omega\! =\! 0.06$, $H_{0}=9.1$, $9.5$, and $9.52$.
		\label{video2}
	}
\end{video}

\begin{figure*}
\includegraphics[width=6.8in]{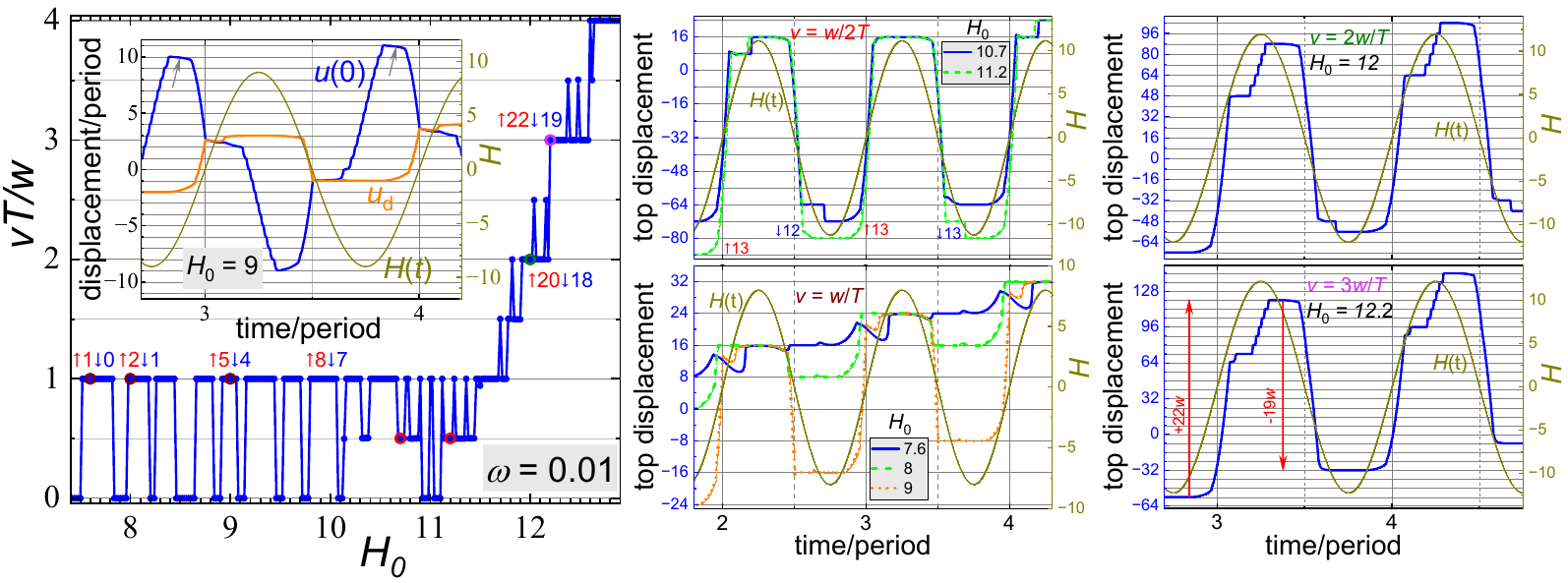}
\cprotect\caption{\cprotect\emph{Left}: The plot of the normalized velocity for $\omega\!=\!0.01$.
\cprotect\emph{Middle and Right} \cprotect\emph{columns}: The top
tip displacement for several field amplitudes
marked on the left velocity plot by the circles corresponding to different
normalized velocities $vT/w$. The directional displacement takes place via the
advance over several periods $n_{f}$ during the increasing-field time interval
and the retreat over several periods $n_{b}$ during the decreasing-field
time interval. Correspondingly, the plateau states at $vT/w\!=\!n_{f}\!-\!n_{b}$
in the left plot are labeled as $\uparrow\!n_{f}\downarrow\!n_{b}$.
The \cprotect\emph{inset} in the left plot shows the time evolution
of both top and bottom displacements for $H_{0}=9$. The gray arrows
point to the small plateaus for the bottom displacement due to the
ratchet potential. %
}
\label{Fig:vNormDisplFr001}
\end{figure*}
\begin{figure}
\includegraphics[width=3.4in]{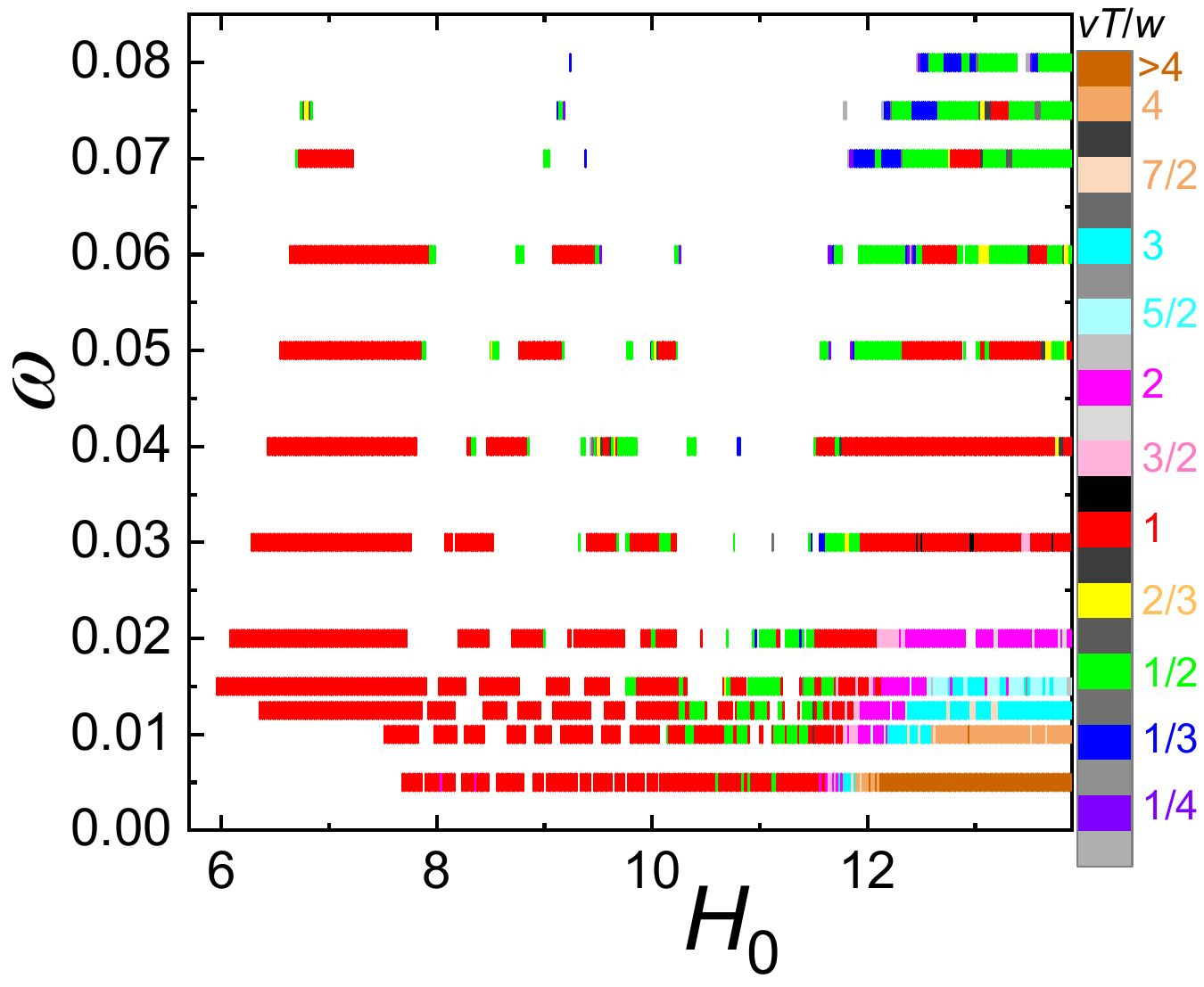}
\caption{Regions of finite DC velocity in the amplitude-frequency plane
color-coded by the reduced velocity $vT/w$. The main plateaus are
coded by colors, while the intermediate states are coded by gray level.}
\label{Fig:H0freqDiagr}
\end{figure}

To illustrate the key features of line dynamics for the simplest regime
with the velocity $w/T$, we show in Fig.~\ref{Fig:DsplTimeCnffr025H8}
a representative plot of the time dependences of the displacements
of top and bottom tips for $\omega\!=\!0.05$ and $H_{0}\!=\!7$.
The images on the right side show snapshots of line configurations
at five time moments marked in the plot. We can observe that the bottom
tip oscillates with a rather large amplitude of $\sim w$, while the
top tip remains pinned in the dip of the ratchet profile most of the
time and snaps to the next dip during a relatively short time interval
(the images 2--4 on the right show the intermediate line configurations
during this snap). The snap initiates when the displacement of the
bottom tip is positive meaning that the top tip is pulled uphill by
the line tension. To compare the rich behavior for different dynamic
regimes, we present in Fig.~\ref{Fig:udtfr0_06} the time evolutions
of the top tip for the motion with different reduced velocities $vT/w\!=\!1/n_{T}$
realized at different field amplitudes for $\omega\!=\!0.06$. We
see that for $n_{T}\!>\!1$ the top tip jumps back and forth between
the neighboring dips before advancing into the neighboring dip of the
ratchet profile, see also the animations for several field amplitudes
for $\omega\!=\!0.06$ in the clip \ref{video2}.

With decreasing the frequency, the dynamic response qualitatively
changes, see the right column in Fig.~\ref{Fig:velHSelected}. The
most dramatic new feature absent at higher frequencies is the formation
of dynamic states with velocity larger than $w/T$. For frequencies
smaller than 0.02, the regions with velocities $n_{w}w/T$ systematically
emerge when the amplitude exceeds $\sim11$ with larger $n_{w}$ values
appearing at higher $H_{0}$. The rise of $n_{w}$ becomes steeper
at lower frequencies. In this range, the intermediate zero-velocity
regions vanish and the velocity is finite at all amplitudes.
The overall increase of the absolute velocity with the amplitude roughly
follows the frequency-independent monotonic curve. This behavior,
however, is not monotonic: the velocity has multiple up and down jumps
between two neighboring states. 

To reveal the origin of this behavior, we show in Fig.~\ref{Fig:vNormDisplFr001}
(middle and right columns) the time evolution of the top displacement
for $\omega\!=\!0.01$ and several field amplitudes representing dynamic
states with different normalized velocity $vT/w$. These amplitudes
are marked in the velocity plot on the left. We see that directional
motion of the top tip takes place by its advance over several periods
$n_{f}$ at the end of the negative-field half cycle. The mechanism
of this advance is similar to that illustrated in Fig.~\ref{Fig:DsplTimeCnffr025H8}
at higher frequency. In the middle of the negative half cycle, the bottom
tip has large positive displacement. As a result, the top tip is
pulled forward by the line tension. At low frequencies, the displacement
of the bottom tip can be very large. For example, in the representative
plot shown in the inset in the plot on the left in Fig.~\ref{Fig:vNormDisplFr001} for
$H_{0}=9$, the displacement of the bottom tip with respect to the
top tip reaches 12 ratchet periods. Importantly, a back motion of
the bottom tip is delayed by the ratchet potential, as indicated by
gray arrows in this plot. This delay is closely related to the static
hysteretic behavior demonstrated in the inset of Fig.~\ref{Fig:umaxH}.
It leads to a slow decrease of the line tension force acting on the
top tip in the time region when the opposing force from the screening
current decreases proportionally to the diminishing magnetic field.
The resulting force imbalance facilitates the advance of the top tip
by multiple periods. On the other hand, at the end of the positive-field
half cycle, the top tip may retreat back by one or several periods
$n_{b}$. The mechanism is similar to the forward advance: the bottom
tip has a large negative displacement in the middle of the positive
half cycle and the top tip is pulled back by the line tension. Due
to the ratchet asymmetry, the maximum negative displacement of the
bottom tip with respect to top tip may be smaller than the positive
displacement. As a consequence, $n_{b}$ may be smaller than $n_{f}$.
In this case the velocity is positive and is given by $(n_{f}\!-\!n_{b})w/T$.
In the left plot of Fig.~\ref{Fig:vNormDisplFr001}, we naturally
label some plateau states with these two numbers as $\uparrow\!n_{f}\downarrow\!n_{b}$.
The multiple up and down jumps with increasing $H_{0}$ are due to
the increase of $n_{f}$ and $n_{b}$, respectively. For amplitudes
smaller than $\sim11$ the velocity jumps from $0$ and $w/T$, where
upward jumps are caused by the increase of $n_{f}$ and downward jumps
are caused by the increase of $n_{b}$. At higher amplitudes, $n_{f}$
increases faster with $H_{0}$ than $n_{b}$ leading to an overall increase
of velocity. Near the crossover amplitude $\sim11$, several intermediate
states with $v=w/2T$ are observed in which the magnitudes of downward
jumps alternate between $n_{b}$ and $n_{b}+1$ for even and odd
periods, as illustrated by the upper plot in the middle column of
Fig.~\ref{Fig:vNormDisplFr001}. In general, the persistence of the
magnetic directional drive down to low frequencies is very important
for its practical implementation. 

To illustrate the general behavior of the rectified velocity, we present
in Fig.~\ref{Fig:H0freqDiagr} the amplitude-frequency diagram, color-coded
by the velocity normalized by the ratio $w/T$. This diagram reveals
several interesting features of the system dynamics: 
(i) The most pronounced
plateaus are realized at the velocities $w/T$, $w/2T$, and $w/3T$,
and the transitions to these plateau states are rather abrupt. 
(ii) The plateau at $w/T$ dominates for $\omega\!\leq\!0.07$ but vanishes
at higher frequencies, $\omega\gtrsim0.075$. 
(iii) The onset field amplitude $H_{\mathrm{on}}$ of
finite velocity depends nonmonotonically on the frequency. The minimal
onset field $\sim 5.96$ is realized at the frequency $\omega\sim 0.015$. 
For frequencies lower than $0.01$, $H_{\mathrm{on}}$ saturates at
$\sim 7.68$ which is slightly lower than the positive static instability
field $H_{+}\!\approx\!7.74$.
(iv) At frequencies above 0.03 the velocity does not exceed $w/T$
in the studied amplitude range but regions with $v>w/T$ systematically
appear at lower frequencies at amplitudes higher than $11$. 

\section{Experimental realization: typical scales for different superconducting
	films \label{Sec:ExpParams}}

The most direct experimental verification of the effect we consider
is the detection of a DC voltage induced by an oscillating in-plane
magnetic field in the presence of a finite static $z$-axis field $B_{z}$.
To relate our numerical results to possible implementations in real
superconducting films with ratchet thickness profiles, we evaluate
in Table \ref{TableParam} the real-unit scales for the magnetic
field, frequency, and velocity in Eqs.~\eqref{eq:ReducedH}, \eqref{eq:frUnit},
and \eqref{eq:velUnit} for films of three common superconducting
materials, niobium, niobium nitride, and molybdenum germanium\cite{NoteParam}.
The estimates are made for temperatures close to the transition temperature
where vortex pinning is weak. We estimated the viscosity coefficient
$\eta$ in the units definitions for frequency and velocity from the
refined Bardeen-Stephen formula $\eta\!=\!1.45\Phi_{0}H_{c2}/(\rho_{n}c^{2})$
\citep{Kopnin2001}. The units estimated in the Table can be used to convert the axes scales in Figs.\ \ref{Fig:DsplTimeCnffr025H8} and \ref{Fig:H0freqDiagr} to physical units. Note that the films we consider are in dirty
limit meaning that the superconducting parameters depend on scattering
rate and vary from film to film.

\emph{Niobium} is a superconducting element with the highest transition
temperature of 9.3 K. This makes niobium films a material of choice
for many applications such as SQUIDS, superconducting electronic prototypes,
and quantum circuits. The parameters of Nb films vary in wide limits depending
on thickness and degree of disorder\cite{AndreonePRB1995,GubinPRB2005,DobrovolskiyPhysRevApplied.13.024012,ValerioCuadrosMat2021,JoshiPRA2023,AltananyPRB2024}.
The parameters listed in the Table \ref{TableParam} are typical for
films used in superconducting electronic circuits \cite{TolpygoIEEETAS2015}.
These films are in dirty limit and their transition temperatures are usually lower than that for clean niobium crystals.  
We have to note that, strictly speaking, the vortex-line model employed
in this paper is applicable in the limit $\xi\ll d_{0}\ll\lambda$
and therefore it does not literally describe Nb films in which the
Ginzburg-Landau parameter $\kappa\!=\!\lambda/\xi$ typically does
not exceed $5$. This does not exclude the possibility of driving vortices
with an AC field in Nb films. It only means that finding conditions for
the realization of this effect requires a more realistic model.
\begin{table*}
	\begin{tabular}{|c|c|c|c|c|c|c|c|c|c|c|c|}
		\hline 
		& $T_{c}${[}K{]} & $\rho_{n}${[}$\mu\Omega\cdot$cm{]} & $d_{0}${[}nm{]} & $\xi_{\mathrm{GL}}${[}nm{]} & $\lambda_{\mathrm{GL}}${[}nm{]} & $T_{c}\!-\!T${[}K{]} & $\xi${[}nm{]} & $\lambda${[}nm{]} & $H_{u}${[}G{]} & $\omega_{u}/2\pi${[}GHz{]} & $v_{u}${[}km/s{]}\tabularnewline
		\hline 
		\hline 
		Nb & 9.1 & 7 & 200 & 11 & 56 & 0.2 & 75 & 380 & 41 & 2.8 & 3.6\tabularnewline
		\hline 
		&  &  & 200 &  &  & 0.5 & 46 & 242 & 59 & 4 & 5.5\tabularnewline
		\hline 
		NbN & 16.5 & 65 & 200 & 4.5 & 177 & 1 & 19 & 740 & 97 & 0.12 & 0.145\tabularnewline
		\hline 
		MoGe & 7.5 & 150 & 200 & 4.6 & 320 & 1 & 13 & 907 & 113 & 0.85 & 1.1\tabularnewline
		\hline 
	\end{tabular}\caption{Superconducting parameters and units for the magnetic field, $H_{u}$,
		frequency, $\omega_{u}$, and velocity, $v_{u}$, in Eqs\@.~\eqref{eq:ReducedH},
		\eqref{eq:frUnit}, and \eqref{eq:velUnit} for films of several superconducting
		materials with thickness $200$ nm. }
	\label{TableParam}
\end{table*}

\emph{Niobium nitride} (NbN) is another popular and well-studied superconducting
material with potential for several applications, in particular,
in single-photon optical detectors \cite{Goltsman2003} or prototype
superconducting electronic circuits \cite{TolpygoIEEETAS23}. A key
advantage of NbN is its relatively high transition temperature, up
to 17 K \cite{Wang1996,KamlapureAPL2010,HazraSST2016}. However, achieving
such high $T_{c}$ requires careful optimization of film growth parameters.
The superconducting parameters of NbN films are also sensitive to
growth conditions. The parameters listed in the Table \ref{TableParam}
correspond to films with transition temperatures close to the maximum
values \cite{KamlapureAPL2010,HazraSST2016}. Contrary to niobium,
NbN is a strongly type-II superconductor with $\kappa\sim40$. Therefore,
this material is expected to be well described by the model used in
this paper. In comparison with niobium films, NbN has a somewhat higher
field scale due to a shorter coherence length, and a significantly lower
frequency scale due to its larger London penetration depth and higher
upper critical field.

Amorphous \emph{molybdenum germanium} (Mo-Ge) films exhibit a relatively
high transition temperature -- up to 7.5K for Mo$_{79}$Ge$_{21}$
-- and high structural homogeneity resulting in a very weak vortex
pinning\cite{GartenPhysRevB.51.1318,Mandal2024}. This material is
characterized by a very small coherence length, leading to a high upper
critical field $H_{c2}(0)\approx12$T \cite{GraybealPhysRevB.29.4167,MaccariPhysRevB.107.014509}
and a large London penetration depth, $\lambda(0)>500$nm\cite{RoyPhysRevLett.122.047001,Mandal2024}.
Due to these properties, Mo-Ge films have been widely used as model
systems to study vortex physics \cite{GartenPhysRevB.51.1318,LiangPhysRevB.82.064502,RoyPhysRevLett.122.047001,LyuNatComm2021,MaccariPhysRevB.107.014509}
meaning that this system is a good candidate for detecting directional
motion of vortices induced by an AC magnetic field. Based on the typical
parameters of a 200 nm Mo-Ge film at $T_{c}-T=1K$ listed in the Table
\ref{TableParam}, we find that the magnetic field scale is comparable
to that of NbN, while the frequency scale is approximately seven time
larger but still smaller than that for Nb. 

An experimental challenge following from Fig.\ \ref{Fig:DsplTimeCnffr025H8}
and the field scales $H_{u}$ listed in Table \ref{TableParam} is
the requirement of relatively large magnetic field AC amplitudes at
relatively high frequencies. The numerical simulations are performed
down to the reduced frequency of 0.005. According to the Table \ref{TableParam},
this corresponds to approximately 15~MHz for Nb film at $T_{c}\!-\!T\!=\!0.2$K,
0.6 MHz for NbN film at $T_{c}\!-\!T\!=\!1$K, and 4.25 MHz for MoGe
film at $T_{c}\!-\!T\!=\!1$K. It is important to note, however, that
the effect does not vanish at low frequencies in our simulations.
Therefore, we expect that the directional motion may persist down
to much lower frequencies, where realization of high magnetic field amplitudes
is experimentally more feasible. The frequency scale also can be reduced
by using thicker films.

\section{Summary\protect\label{sec:Summary}}

In summary, we performed molecular-dynamics simulations of a single
vortex line in a superconducting film with an asymmetric ratchet pinning
potential in the form of a sawtooth thickness profile. We demonstrated
that the line may be directionally advanced by the oscillating magnetic
field applied parallel to the film. We quantitatively characterized
dynamic behavior for different amplitudes and frequencies of the AC
drive for a representative realization of the ratchet profile and
revealed several qualitatively different dynamic regimes. These results
may be useful for the elaboration of efficient techniques to remove undesired
trapped magnetic flux from contactless superconducting circuit elements.

\begin{acknowledgments}
The author would like to thank Leonardo Cadorim, Milorad Milo{\v s}evi{\'c},
and Boldizsar Janko for valuable discussions and Ulrich Welp for critical
reading of the manuscript and useful comments. This research was sponsored
by the Army Research Office and was accomplished under Grant Number
W911NF-24-1-0145. The views and conclusions contained in this document
are those of the authors and should not be interpreted as representing
the official policies, either expressed or implied, of the Army Research
Office or the U.S. Government. The U.S. Government is authorized to
reproduce and distribute reprints for Government purposes notwithstanding
any copyright notation herein.
\end{acknowledgments}

\bibliography{VortexLineRatchetACfield}

\begin{thebibliography}{57}
\bibitem [{Not()}]{NoteParam}%
  \BibitemOpen
  \href@noop {} {}\bibinfo {note} {The parameters $\xi_{\mathrm{GL}}$ and
  $\lambda_{\mathrm{GL}}$ inthe Table \ref{TableParam} are the Ginzburg-Landau
  lengths which determine the temperature dependence of the coherence length
  and Londonpenetration depth near the transition temperature $T_{c}$,
  $\xi(T)\!=\!\xi_{\mathrm{GL}}/\sqrt{1\!-\!T/T_{c}}$and
  $\lambda(T)\!=\!\lambda_{\mathrm{GL}}/\sqrt{1\!-\!T/T_{c}}$.}\BibitemShut
  {Stop}%
\end{thebibliography}

\appendix

\section{Numerical procedure\protect\label{app:NumProc}}

In this appendix we briefly describe the numerical procedure to solve
the dynamic equation in Eq.~\eqref{eq:VLdynRed}. We discretize the
continuous function $u(z)$ as an array $u_{n}$ with the $z$ step
equal to $s$. The numerical mesh is illustrated in Fig\@.~\ref{fig:VortLineRatchet}.
The challenge is that the effective size of the array $N+1$ is not
fixed but depends on the displacement at the top. It has to satisfy
the inequalities 
\begin{equation}
s(N-\tfrac{1}{2})<d(u_{N}),\,s(N+\tfrac{1}{2})>d(u_{N+1}).\label{eq:NCond}
\end{equation}
We discretize the derivative in the line tension term in Eq.~\eqref{eq:VLdyn}
as
\begin{widetext}
\begin{align*}
\left[\frac{\mathrm{d}}{\mathrm{d}z}\frac{1}{\sqrt{1\!+\!\left(\frac{\mathrm{d}u}{\mathrm{d}z}\right)^{2}}}\frac{\mathrm{d}u}{\mathrm{d}z}\right]_{n} & =\frac{1}{s}\left\{ \left[\frac{1}{\sqrt{1\!+\!\left(\frac{\mathrm{d}u}{\mathrm{d}z}\right)^{2}}}\frac{\mathrm{d}u}{\mathrm{d}z}\right]_{n+1/2}\!\!-\left[\frac{1}{\sqrt{1\!+\!\left(\frac{\mathrm{d}u}{\mathrm{d}z}\right)^{2}}}\frac{\mathrm{d}u}{\mathrm{d}z}\right]_{n-1/2}\right\} \\
=\frac{1}{s^{2}} & \left\{ \frac{u_{n+1}-u_{n}}{\sqrt{1\!+\!\left(\frac{u_{n+1}-u_{n}}{s}\right)^{2}}}-\frac{u_{n}-u_{n-1}}{\sqrt{1\!+\!\left(\frac{u_{n}-u_{n-1}}{s}\right)^{2}}}\right\} 
\end{align*}
Therefore, the equation for $u_{n}(t)$ is
\begin{equation}
	\frac{\mathrm{d}u_{n}}{\mathrm{d}t}  =\frac{1}{s^{2}}\left[\frac{u_{n+1}-u_{n}}{\sqrt{1\,+\,\left(\frac{u_{n+1}-u_{n}}{s}\right)^{2}}}-\frac{u_{n}-u_{n-1}}{\sqrt{1\,+\,\left(\frac{u_{n}-u_{n-1}}{s}\right)^{2}}}\right]
	 +\,\tilde{H}(t)\left[s(n-\frac{1}{2})\,-\frac{d(u_{n})}{2}\right]\label{eq:VLdynDiscr}
\end{equation}
with $n=1,\ldots,N$. 

For the bottom tip, we update $u_{0}$ by setting $u_{0}=u_{1}$.
The procedure to update the top-tip displacement is:
\begin{itemize}
\item Update the tip point $u_{N}$ using Eq.~\eqref{eq:VLdynDiscr}
\item Update outside reference point $u_{N+1}$ from BC in Eq\@.~\eqref{eq:BCd}
\item Update $N$ if one of the conditions in Eq.~\eqref{eq:NCond} fails
\begin{itemize}
\item If $s(N-\tfrac{1}{2})>d(u_{N})$, tip point moved outside, change
$N\rightarrow N-1$
\item If $s(N+\tfrac{1}{2})<d(u_{N+1})$, outside reference point moved
inside, change$N\rightarrow N+1$, and define new $u_{N+1}$ from BC
\end{itemize}
\end{itemize}
To update $u_{n}$ in one time step, we use an implicit procedure.
Defining $u_{n}=u_{n}(t)$ and $u_{n}^{+}=u_{n}(t+h_{t})$, we obtain
from Eq.~\eqref{eq:VLdynDiscr}
\begin{align}
u_{n}^{+}-u_{n} & =\frac{h_{t}}{2s^{2}}\left[c_{n}^{+}\left(u_{n+1}^{+}\!-\!u_{n}^{+}\right)\!-\!c_{n-1}^{+}\left(u_{n}^{+}\!-\!u_{n-1}^{+}\right)\!+\!c_{n}\left(u_{n+1}\!-\!u_{n}\right)\!-\!c_{n-1}\left(u_{n}\!-\!u_{n-1}\right)\right]\nonumber \\
 & +\,h_{t}\left[\frac{\tilde{H}^{+}+\tilde{H}}{2}s(n-\tfrac{1}{2})\,-\frac{\tilde{H}^{+}d(u_{n}^{+})+\tilde{H}d(u_{n})}{4}\right]\label{eq:VLdynTimeStep}
\end{align}
\end{widetext}
with 
\[
c_{n}=\frac{1}{\sqrt{1\,+\,\left(\frac{u_{n+1}-u_{n}}{s}\right)^{2}}}.
\]
This is nonlinear equations for $u_{n}^{+}$. To reduce them to linear
equations, we have to make approximations $c_{n}^{+}\approx c_{n}$,
$d(u_{n}^{+})\approx d(u_{n})$. We also introduce notation $r_{t}=h_{t}/s^{2}$.
This gives
\begin{align}
 & \left[1\!+\frac{r_{t}}{2}\left(c_{n}\!+\!c_{n-1}\right)\right]u_{n}^{+}-\frac{r_{t}}{2}\left[c_{n}u_{n+1}^{+}+c_{n-1}u_{n-1}^{+}\right]\nonumber \\
= & \left[1-\frac{r_{t}}{2}\left(c_{n}+c_{n-1}\right)\right]u_{n}+\frac{r_{t}}{2}\left[c_{n}u_{n+1}+c_{n-1}u_{n-1}\right]\nonumber \\
 & +\,h_{t}\frac{\tilde{H}^{+}+\tilde{H}}{2}\left[s(n-\tfrac{1}{2})-\frac{d(u_{n})}{2}\right],\label{eq:VLdynTimeStepLinear}
\end{align}
This is a tridiagonal linear system 
\begin{align*}
A_{n}u_{n-1}^{+}+B_{n}u_{n}^{+}+C_{n}u_{n+1}^{+} & =F_{n}
\end{align*}
with the coefficients and the right-hand side
\begin{align*}
A_{n}= & -\frac{r_{t}}{2}c_{n-1},\,B_{n}=\left[1\!+\frac{r_{t}}{2}\left(c_{n}\!+\!c_{n-1}\right)\right],\\
C_{n}= & -\frac{r_{t}}{2}c_{n},\\
F_{n}=\! & \left[1\!-\frac{r_{t}}{2}\left(c_{n}\!+\!c_{n-1}\right)\right]u_{n}\!+\frac{r_{t}}{2}\left[c_{n}u_{n+1}\!+\!c_{n-1}u_{n-1}\right]\\
+ & \frac{\tilde{H}^{+}\!+\!\tilde{H}}{2}\left[s(n\!-\tfrac{1}{2})\!-\frac{d(u_{n})}{2}\right].
\end{align*}
This tridiagonal system can be solved using Thomas algorithm allowing
to present the solution in the form
\begin{align}
u_{n}^{+} & =\alpha_{n+1}u_{n+1}^{+}+\beta_{n+1},\label{eq:Progonka}
\end{align}
where the coefficient $\alpha_{n}$ and $\beta_{n}$ can be found
iteratively as 
\begin{align}
\alpha_{n+1} & =-\frac{C_{n}}{A_{n}\alpha_{n}+B_{n}},\,\beta_{n+1}=\frac{F_{n}-A_{n}\beta_{n}}{A_{n}\alpha_{n}+B_{n}},\label{eq:ProgCoef}\\
\alpha_{1} & =1,\,\beta_{1}=0.\nonumber 
\end{align}
The iteration in Eq\@.~\eqref{eq:Progonka} starts from the top
value $u_{N+1}^{+}$, which has to be found from the boundary condition
\begin{align*}
\frac{u_{N+1}^{+}-u_{N}^{+}}{s} & =-\frac{d(u_{N}+du)-d(u_{N})}{du},
\end{align*}
which together with Eq.~\eqref{eq:Progonka} yields
\begin{align}
u_{N+1}^{+} & =\frac{\beta_{N+1}-s\frac{d(u_{N}+du)-d(u_{N})}{du}}{1-\alpha_{N+1}}.\label{eq:uN+1}
\end{align}
Therefore, the time advance procedure is: (i) find the Thomas algorithm
coefficients using Eq.~\eqref{eq:ProgCoef}, (ii) find the new top
displacement using Eq.~\eqref{eq:uN+1}, and (iii) compute all new
displacements recursively using Eq.~\eqref{eq:Progonka}. After the
full time step is completed, we check the conditions in Eq.~\eqref{eq:NCond}
and, if necessary, update $N$.
\end{document}